# Genetic Diversity Assessment of German Chamomile Populations Based on Phenotypic Detection and SSR Molecular Markers in Bushehr Area

Seyyedeh Mahdiyeh Hashempour[1], Mohammad Modarresi[1]*, Mojtaba Ghasemi[2], Ivan Salamon[3]*

[1] Department of Plant Genetics and Production Engineering, Faculty of Agriculture, Persian Gulf University, Bushehr 7516913817, Iran

[2] Department of Biotechnology, The Persian Gulf Research Institute, Persian Gulf University, Bushehr 7516913817, Iran

[3] Department of Ecology, Faculty of Humanities and Natural Sciences, University of Prešov, 08001 Prešov, Slovakia

* Authors to whom correspondence should be addressed.

## Abstract

German chamomile (*Matricaria chamomilla*) is an herbaceous and annual plant from the Asteraceae family. Chamomile has been known as an important medicinal plant to use its extract and essential oil in the pharmaceutical, cosmetic, health industries, perfumery, and food seasonings. The present study aimed to investigate the genetic diversity in 42 German chamomile populations using morphological, biochemical, and ssr molecular markers, in both field and laboratory sections. The field part was conducted at the Faculty of Agriculture and Natural Resources of the Persian Gulf University and the molecular markers laboratory part was conducted on leaf samples harvested from the field at the Persian Gulf Research Institute. Different populations of German chamomile collected from different regions were as statistical populations. Phenological traits, days to emergence and days to flowering; by calculating the number of days to 70% emergence and days to 50% flowering, respectively; plant height, diameter of flowers and receptacles, fresh and dry weight in each fold, essential oil content, percentage of camazulene in essential oil and chlorophyll a and b were measured. DNA extraction was done by CTAB method and DNA quantity and quality were checked by spectrophotometer and gel electrophoresis. 5 target populations were genetically evaluated with the help of 6 SSR markers. According to the obtained results, Jam 1 and Shahijan populations had the most effective substances (essential oil and chamazulene). The correlation analysis revealed a positive and significant correlation of 59% between the percentage of chamazulene and the percentage of essential oil. Chamazulene percentage and fresh weight as the most important traits were entered into the regression model step by step. These traits were found to explain 0.43% of the changes in the data. These findings have significant implications for future research aimed at identifying the most effective populations for essential oil and chamazulene production. The cluster analysis divided the genotypes into five groups. The second group was the most important, and Jam1 and Shahijan genotypes were in this group. The results of the molecular analysis showed that the Seho Sermak-Dashti population had the most effective allele, and this population was superior to other populations in terms of Shannon's index and nei diversity coefficient. According to the diagram, analysis into main coordinates showed that the genotypes are scattered on the surface of the diagram and this indicates the appropriate diversity of the studied genotypes. As a result, it can be stated that the grouping of phenotypic and molecular data was very consistent with each other.

# Introduction

Chamomile is one of the plants that is widely used in various pharmaceutical, food, and ornamental industries today. Also, it has many medicinal properties, among which we can mention the effects of sedative, antispasmodic, stimulating the body and strengthening the anti-allergic defense system [1,2]. Iran has valuable genetic reserves of medicinal plants, but in recent decades, most research and efforts have focused on preserving and utilizing agricultural and garden plant germplasm, with little attention given to medicinal and ornamental plants. Multi-breeding programs have made great advancements in enhancing and expanding the production of agricultural products by creating and introducing new and improved varieties, breeds, and genotypes [19]. Therefore, understanding the pattern of genetic diversity of species is crucial for preserving the species, preventing genetic drift, and building collections of genetic resources. Additionally, studying the genetic diversity within populations of a species is essential for the success of future breeding programs [12,16]. Genetic diversity enables plant breeders to identify or produce populations of plants with high performance and greater adaptability to environmental changes. This is done by using diversity through selecting suitable populations and using them in breeding programs [9]. The differences in DNA sequences between living organisms can also be reflected in their physical appearance, making them useful as markers to distinguish between individuals. Morphological traits that are mainly controlled by one gene can also be used as genetic markers. The existence of genetic diversity and the evaluation of important traits enables plant breeders to identify, select, or produce high-yielding populations or genotypes that are more adaptable to environmental changes by using the available diversity [3][6]. Molecular markers are another way used to identify genetic variation. In the present research, the SSR molecular marker was used to study the genetic affinity of 42 German chamomile populations cultivated in Iran. As a result, the use of SSR microsatellites provides a successful approach for marker-assisted selection in plant breeding programs [17]. The development and use of simple repetitive sequences of SSR markers associated with specific traits allows the selection of desirable genotypes instead of phenotypes and hence can accelerate plant breeding programs [13]. This method can speed up the process of breeding plants with desired traits. SSR markers have been found to be very useful in studying plant genetics due to their high variability among different plant populations. They have been widely used for almost three decades and are expected to continue being an important tool for studying genetic diversity in plant species in the future[14]. Researchers investigated the genetic diversity of *Ambrosia artemisiifolia* L. based on 13 SSR primers and 13 EST-SSR primers. The findings revealed significant genetic diversity between populations in North America and Europe. They also stated that The SSR marker has a high efficiency in investigating the genetic diversity between the populations of the studied species. The researchers also stated that the SSR marker was highly effective in assessing the genetic differences among the populations of the studied species. Oken et al. (2013) in the study of the genetic diversity of 15 wild chamomile genotypes with ISSR markers reported that out of 48 bands produced by primers, 41 were polymorphic bands (85.4%). in this research, the genetic similarity coefficient was between 0.5 - 0.52, and the average genetic similarity was calculated to be 0.653. Additionally, through cluster analysis, 15 studied populations were divided into three main groups based on their genetic similarities [15].

Ahmadi et al. (2014) investigated the genetic diversity of 23 populations of German chamomile using the ISSR marker and they found that the populations in the central regions of Iran exhibited the highest genetic diversity, with a value of 0.25. Also, the main coordinates confirmed the results of the cluster analysis, and a total of 193 bands were produced from the 10 selected primers, which had an average of 19.3 polymorphic bands. [10]. In 2012, a study was conducted by Ahmadi on the genetic diversity of German chamomile in 21 Iranian populations and two European populations. The study employed ISSR molecular markers and morphological characteristics and utilized 10 ISSR primers. The study produced a total of 208 bands with a clear pattern, out of which 193 bands (92.78 percent) were polymorphic. Analysis of molecular variance revealed that the diversity within the group (58.98%) was higher than the diversity between the groups (41.02%) [1]. This suggests that there is significant genetic diversity within the populations studied, with more variation found within groups than between them. The aim of this investigation was to evaluate the genetic diversity of chamomile populations collected from different places based on the data obtained from morphological traits and molecular markers, as well as evaluating the usefulness and efficiency of the investigated primers in creating polymorphism.

# Material and methods

This study was conducted during the agricultural year of 2019 at the research farm of the Faculty of Agriculture and Natural Resources of the Persian Gulf University, situated near Borazjan. The tested treatments include 42 populations of German chamomile collected from different regions of Bushehr province, which were cultivated in a randomized complete block design with three replications. Before planting, the desired chemical fertilizers (including sulfur, urea, potassium, and phosphate) were added to the field soil during the preparation of the land. Experimental plots were then set up in blocks with a distance of one meter between plots and three meters between blocks.

For optimal germination conditions, direct and completely superficial seed cultivation is employed, as the seeds require surface planting, light, and a temperature of 20 degrees Celsius for germination. This method ensures an ideal environment for germination, thereby facilitating the growth of chamomile populations.

Table 1- Chamomile medicinal plant populations evaluated in the research farm

| Number | Treatment | Number | Treatment |
|---|---|---|---|
| 1 | Tange Faryab | 22 | Tange eram |
| 2 | Jam 2fd | 23 | Noken abdan |
| 3 | Izeh khozestan | 24 | Jam 1 |
| 4 | Dehrod Sofla | 25 | Ghale toil khozestan |
| 5 | Shahijan | 26 | Tal bardi |
| 6 | lalab | 27 | Lavar sharghi |
| 7 | Haris Village | 28 | Dehrod sofla |
| 8 | sahosarmak | 29 | unknown |
| 9 | Urea Firozabad | 30 | Pol dokhtari |
| 10 | Ghalat nilo | 31 | Jam v riz |
| 11 | Malek norouz garden | 32 | Zeraei Tehran |
| 12 | Baharestan jam | 33 | Bona |
| 13 | Asasd abad behbabhn | 34 | Hamedan 1 |
| 14 | Loran khozestan | 35 | Harami anari |
| 15 | Eslam abad jam | 36 | Kori manzar |
| 16 | pashto | 37 | Kalameh tangezard village |
| 17 | Saho sarmak 2 | 38 | Selected Bona Agriculture and Natural Resources faculty borazjan |
| 18 | Anarestan galobardakan | 39 | Alle khorshid khozestan |
| 19 | jamklati | 40 | Galobardekan (pankareh) |
| 20 | kazeron | | Tekye garde khozestan |
| 21 | Khoviz mountain | | German chamomile with an unnatural receptacles |

To assess the specific characteristics of the plant, various traits were measured during its growth stages. These included phenological traits such as the number of days to greening, and the number of days to flowering, as well as morphological traits such as plant height, stem diameter, petal diameter, and fresh and dry weight of flowers. Additionally, phytochemical traits including the percentage of essential oil, the percentage of essential oil and chamazulene, chlorophyll a, chlorophyll b, and total chlorophyll were measured. To measure the phytochemical characteristics, chamomile essential oil was first extracted from a certain amount of dry flowers, and then a spectrophotometer was used to measure the amount of chamazulene. In order to determine the yield of chamomile essential oil and active substance, a mixture of different crushed flowers was prepared for extraction, using 30 grams of dried flowers. To extract the essential oil, a mixed sample of crushed dried flowers was poured into a 1000 ml flask with 600 ml of distilled water and the essential oil was extracted using a Clevenger device using water distillation for three hours. Arnon's (1967) method was used to measure the content of chlorophyll a, b, and total chlorophyll. In this method, 0.5 g of fresh chamomile leaf tissue was gradually ground using 10 ml of 80% acetone in a Chinese mortar (Figure 4-3). The obtained extract was centrifuged for 10 minutes at a speed of 6000 rpm, and then the light absorption of the supernatant solution was read by a spectrophotometer at 663 and 645 nm wavelengths, and the amount of chlorophylls a and b were calculated according to the following equation.

$$\text{Chlorophyll a (mg/g)} = [12.7\,(A_{663}) - 2.69(A_{645})] \times V/(1000\times W)$$

$$\text{Chlorophyll b (mg/g)} = [22.9\,(A_{663}) - 4.68(A_{645})] \times V/(1000\times W)$$

$$\text{Total Chlorophyll (mg/g)} = [20.2\,(A_{663}) + 8.02(A_{645})] \times V/(1000\times W)$$

V = volume of acetone
W = fresh weight of the leaf sample (g)
$A_{663}$ and $A_{645}$ = wavelength (nm)

The molecular marker used in the present study for the genotypic evaluation of the populations is SSR and the genotypes were classified based on the obtained genotypic data. The method of conducting molecular tests, includes leaf sampling, DNA extraction, determining the quantity and quality of DNA, preparing and performing the desired marker steps, performing acrylamide electrophoresis, recording marker data, and analyzing the resulting data, respectively. Out of the 42 target populations, 5 groups were chosen based on their superior morphological and phytochemical traits to perform molecular work in the laboratory. DNA was extracted using a manual method called CTAB. The desired marker was SSR. 6 specific polymorph primer pairs were used to record the genotypic data of these selected populations. Analysis of morphological and biochemical data, which included univariate and multivariate statistics, was performed using SAS9.4, Excel2013, and SPSS statistical software. In this research, to analyze the results of the SSR molecular marker, the bands obtained from electrophoresis gel were scored and the data were entered using Excel software. Also, the method of decomposition into main coordinates was done by Gen AIExe software.

## Results and discussion

To compare the populations being studied, an analysis of variance was conducted after ensuring the normality of data distribution. The results of the studied traits are presented in Table 2. The analysis of variance results indicated significant variability at a one percent probability level for all traits except for the number of days to greening among the populations being evaluated. This suggests that there is a suitable genetic diversity among the evaluated populations. This issue indicates the existence of suitable genetic diversity between different chamomile medicinal plant populations. The results of the mean comparison, correlation, stepwise regression, causality analysis, and cluster analysis are presented in separate tables below (Table 2). The analysis of the average traits revealed that Bona population (with an average of 74.54 cm), Dehroud Sofla population (with an average of 75.16 cm), Shahijan population of Fars province (with an average of 72.68 cm), and Harami Anari population (with an average of 72.38 cm) had the highest plant height compared to other populations which there was no significant difference between these populations in terms of this attribute. The lowest height related to the population of Pol Dokhtari was 49.18 cm. These results were consistent with the results of Ahmadi et al. (2017). Also, the results showed that the population of Qalat Nilo (with an average of 21.56 mm), Shahijan (with an average of 21.43 mm), Hamedan 1 (with an average of 20.58 mm) and German Chamomile with unnatural receptacles (with an average of 20.50 mm) ) accounted for the largest flower diameter. Urea population had the lowest mean (30.4 mm). In the case of flower diameter, in general, a high range of changes was observed, and their difference was significant at the statistical level of 1%. The main reason for this significant difference is the existence of the Urea population without petals along with the studied populations with petals. However, there are significant differences between the populations with petals. If the climatic conditions of the populations are carefully examined, perhaps a relationship can be found between the population's climate and this characteristic. The receptacle diameter trait was found to be significantly larger in Hamedan population 1 (7.67 mm) and showed a statistically significant difference with other populations, followed by Tang Eram population (with an average of 6.94 mm) and Shahijan (with an average of 6.90 mm), which had the largest receptacles diameter. In the studies of Zeinali et al. (2009), which were conducted on fourteen German chamomile populations, a significant difference was observed for flower diameter at the five percent probability level, and the range of flower diameter changes was reported between 9.47 - 8.37 mm. The Hamadan 1 population had a significantly larger receptacle diameter of 8.59 mm compared to other populations, according to Darvishi et al.'s study (2017). It showed a statistically significant difference with other populations. On the other hand, the populations of Urea, Dehroud Soflai, and Globardakan of Anarestan had a very small receptacle diameter of 6.5 mm, which was significantly different from most populations. The populations of German chamomile that had the highest amount of fresh and dry weight were Qalat Nilo (with an average of 182.17 and 81.25 grams), Shahijan (with an average of 180.20 and 98.81 grams), Jam 1 (with an average of 178.71 and 21.96 grams), and Khoviz mountains (with an average of 173.87 grams). However, the Urea population was found to be characterized by small flowers and a smaller  capitoldiameter when compared to other populations. It also had the lowest average fresh weight. In a study conducted by Ahmadi et al. (2016) on the genetic diversity of German chamomile, it was observed that the Urea population had the lowest fresh and dry weight when compared to other populations, as shown in Table 3. In the analysis of the essential oil values of the investigated populations, it was found that the highest amount of essential oil is related to Jam 1 population (with an average of 0.40), Shahijan (with an average of 0.38), Chamklati (0.39 percent), Baharestan Jam (0.38), which had a statistically significant difference with other populations. The range of essential oil content of these populations was found to be between 0.11 and 0.12, which is the lowest value recorded for this trait. It is important to note that the amount of essential oil in plants is influenced by both genetic and environmental factors, like most other plant traits. In this research, since all genotypes were grown in the same environment and under the same conditions, the difference in the amount of essential oil can be partially attributed to the genotypic difference; Therefore, it is expected that populations with a higher amount of dry flowers have more essential oil and chamazulene , which is especially true for Jam 1 and Shahijan populations (Table 4). Seluki et al. (2008) investigated 20 chamomile populations and reported the essential oil content of the studied masses between 0.1-0.8%. Tavini et al. (2002) in their studies on 15 populations of German chamomile collected from Europe, reported a high diversity for yield and quality traits, also the essential oil of the studied populations was reported between 0.89-0.29% [20]. In 2004, Salamon conducted a study on 29 habitats located in Slovakia. The study reported the presence of chamomile essential oil in the range of 0.5 to 0.91%. Furthermore, a significant correlation was observed between the amount of essential oil and the geographic distribution of germplasm in Slovakia. [18]. The comparison of averages for the chamazulene percentage showed that the chamazulene percentage of populations was in a wide and significant range between 10.37-1.88%. The population of Shahijan (10.37), Jam 1 (with an average of 9.90) and Qalat Nilo (with an average of 9.91) had the highest amount of chamazulene. The populations of Bushehr province and Kazeroon also showed a high percentage of chamazulene, with most of them recording above 8%. These populations exhibited the highest values of chamazulene, which is an important biochemical trait (Table 4). On the other hand, the populations of Urea (1.88%) and Tehran Zraai (1.23%) had the lowest chamazulene percentage values, which were significantly different from the other populations. These findings were consistent with those of Ahmadi et al. (2016). The most positive and significant correlation (Table 5) was obtained between flower diameter and receptacle diameter (62%), fresh weight and dry weight (88%), chamazulene and percentage of essential oil (59%), chlorophyll a and chlorophyll b (80%), chlorophyll a and total chlorophyll (79%), chlorophyll b and total chlorophyll (90%), number of days to flowering and diameter of receptacle (40%) and days to greening and rose to flowering (47%) (Table 5). Pirkhezari et al. (2008) showed a positive and significant correlation between yield and number of flowers per plant, yield weight of 100 flowers, flower diameter and number of tabular flowers, flower diameter and yield, plant height and leaf length, yield and number of tabular flowers. The analysis of correlation coefficients in the study of Golparvar et al. (2008) showed that the number of flowers per plant, yield of fresh flowers per plant, days to budding, days to 50% flowering, days to 100% flowering and the number of flowering stems have a direct and significant relationship with dry yield in the plant. A study conducted by Mohammadi and colleagues in 2013 revealed that certain traits, such as the number of flowers per plant, fresh weight, and dry weight of flowers in chamomile plants, are significantly correlated with flower yield. These traits can be used as good indicators to select plants with higher yields. The study also demonstrated a positive and significant correlation between phenological traits. This indicates that by examining the number of days until flowering, we can make an early and suitable estimate of the length of the growth period of plants. These findings can be useful in breeding programs to increase or decrease the length of the growth period [9][11]. The findings of this study offer valuable insights into optimizing breeding programs for higher crop yields and improving overall productivity in the agricultural industry. They also discovered that certain traits, such as flower diameter and specific chemical components like chamazulene, play a significant role in determining the percentage of essential oil produced by the plant. By understanding these relationships, breeders can make more informed decisions to improve essential oil production in chamomile plants. Stepwise regression was conducted in addition to the correlation analysis of traits. Since essential oil is a significant measure of chamomile medicinal plant production, a stepwise regression was performed using the percentage of essential oil as a dependent variable. Chamazulene percentage and wet weight were selected as the two most important traits and were entered into the model stepwise. These traits justified a total of 0.43% of data changes and remained in the model (as shown in Table 6). The researchers found that traits such as chamazulene percentage and wet weight were important factors affecting the percentage of essential oil. A study on the genetic diversity of German chamomile was conducted using stepwise regression based on the percentage of essential oil as the dependent variable by Ahmadi et al. (2017). Four traits, including flower diameter, height, fresh flower weight, and dry flower weight, were entered into the model as the most important components of yield, which justified 59% of the data. The diameter of the flower was the first trait included in the stepwise regression model. It explained 28% of the variation in the total percentage of essential oil. This suggests that this trait plays a crucial role in increasing the percentage of essential oil production. It also highlights the potential of this trait for use in breeding programs. [1]. While correlations between traits were examined, the study also emphasized the importance of causality analysis to understand the direct effects of specific traits on essential oil production. The results showed that the chamazulene trait had the highest direct effect, with a value of 0.680, indicating its significant influence on essential oil percentage. This finding was consistent with the correlation data, which showed a strong relationship (0.560) between chamazulene and essential oil percentage. Overall, the study suggests that understanding the causal relationships between traits is crucial for optimizing essential oil production in chamomile plants. According to the research findings, wet weight had the most significant negative effect, which is consistent with the correlation data and is statistically significant. On the other hand, the highest positive indirect effect was linked with fresh weight through chamazulene (0.120), while the most significant negative indirect effect was related to chamazulene through fresh weight (-0.343). Therefore, considering that the direct and indirect effects of traits are in line with correlation coefficients, it can be concluded that correlation coefficients can accurately explain the real relationship between two variables, and direct selection of essential oil percentages through these traits can be useful. These findings are presented in Table 7. Ahmadi et al. (2016) investigated the genetic diversity of German chamomile. They reported that in the causality analysis for the percentage of essential oil as a dependent variable, the largest direct effect was related to the fresh weight of the flower (1.09). The largest indirect effect was related to the fresh weight of the flower through dry weight (1/06) [1]. Darvishi et al. (2017) in a study of 21 German chamomile populations stated that the characteristics of chamazulene and the dry weight of flowers had the most direct effect on the percentage of essential oil. They also stated that the direct and indirect effects of traits are consistent with the correlation coefficients and direct selection for essential oil percentage through these traits can be useful [5]. Ahmadi et al. (2016) conducted a study on the genetic diversity of German chamomile. They found that the fresh weight of the flower had the largest direct effect on the percentage of essential oil as a dependent variable (1.09), while the largest indirect effect was related to the fresh weight of the flower through dry weight (1/06) [1]. In another study by Darvishi et al. (2017) on 21 German chamomile populations, it was reported that characteristics such as chamazulene and dry weight of flowers had the most direct effect on the percentage of essential oil. The researchers also suggested that direct selection for essential oil percentage through these traits could be useful, as the direct and indirect effects of traits were consistent with the correlation coefficients [5]. Using component decomposition, 12 components were obtained, and the first five components with eigenvalues greater than one were selected (Table). These five components in total explained about 83.40% of the phenotypic variation (table). In the first component, the characteristics of chlorophyll a, chlorophyll b and total chlorophyll were more important and for this reason, it was called the photosynthetic capacity component. The second component also had a strong positive relationship with 18.50 percent of variations with height, diameter of flower, flower diameter, fresh weight, and dry weight of flower. Therefore, this component was called the component of morphological characteristics. The percentage of essential oil and chamazulene in the third component, which justified 11.15% of the variations, were more important and can be named as the pharmaceutical quality component. The characteristics of fresh weight, dry weight, chamazulene percentage, and essential oil percentage in the fifth component had the highest positive and high coefficients in the explanation of this component, which led to it being named the medicinal grade component. This component includes 9.09 percent of the variation. In the fifth component, the characteristics of days to greening, days to flowering and height had a strong relationship in a positive direction, so this component can be named as the component of the maturity period. This component justified 16% of the variation changes (Table 8). Mohammadi et al. (2013) investigated the genetic diversity of German chamomile populations. The study involved the analysis of 14 components, from which three key components were identified and accounted for 82.37% of the total diversity. In the first component, the characteristics of harvest index, fresh and dry weight of 50 flowers, flower yield, plant height, flower height and number of flowers per plant had positive and high factor loading. In the second component, the number of days to blooming, the number of days to the first flowering, the number of days to 50% flowering, and the number of days to full flowering, and in the third component the traits of flower diameter, number of flowering substems and biological yield had the highest positive coefficient [9]. By performing cluster analysis, 42 studied populations were divided into five groups. The first cluster includes Tang Faryab populations, elective of the Borazjan Faculty of Agriculture, Lalab Khuzestan, Malek Nowruz garden, Sahosarmak 2, which had the highest amount of chlorophyll a, chlorophyll b, total chlorophyll and the number of days until flowering, and on the other hand, they recorded the lowest height. In the second cluster, the populations of Jam 2, Qalat Nilo, Khoviz mountains, Pashto, Shahijan, Sahosarmak Dashti region, Jam 1, and Tekyeh Gard of Khuzestan were placed, which had the highest amount of capitol diameter, flower diameter, fresh weight, dry weight of flower, percentage of essential oil and chamazulene amount. This group was the best group for all important traits.

The populations of Urea, Dehroud Sofla, and Kori Manzar were placed in the third cluster. These populations showed the lowest averages for traits such as days to flowering, fresh weight of flowers, diameter of flowers, dry weight of flowers, chlorophyll a, chlorophyll b, percentage of essential oil and total chlorophyll. Therefore, it can be concluded that these populations have the lowest values for most of the traits and are not suitable for the selection of diversity. The fifth group had only one population, which was the German chamomile with an abnormal flower head. This population had the highest values of chlorophyll a, chlorophyll b, and total chlorophyll, flower diameter, capitol diameter, and number of days to flowering, and had low flower dry weight. The third group was the largest cluster of this cluster, which included many populations with high diversity. The results of the experiment showed that the second and fourth clusters had the largest genetic distance between them. These two populations differed from each other in most of the traits and had a maximum genetic distance, making them potential candidates for breeding programs. Due to their high heritability and desired traits, they could be considered as potential parents for breeding programs. Also, the creation of genetic populations with higher value and the occurrence of high heterosis is not far from expected. It is also expected that the maximum genetic diversity will be created by crossing the populations of these two clusters and its results will be used as raw materials for cultivar improvement (Figure 1). According to grouping and estimating the average traits of populations in each cluster, suitable parents can be selected for cross-breeding programs. Also, based on the grouping, the genetic masses related to a region were placed in separate groups. This shows that genetic diversity does not follow geographical diversity. Seloki et al. (2008) also investigated 20 genetic masses collected from different regions of Iran and Europe and reported that no clear patterns were observed between genetic and geographical diversity. Genetic diversity not following geographical diversity may be due to the relative similarity of the environmental and ecological conditions of the cities where the samples were collected, as well as the transfer of germplasm between different regions. The study also found that genetic diversity does not necessarily follow geographical diversity, as seen in the grouping of genetic masses from different regions.

## Analysis of Molecular Variance (AMOVA)

According to the analysis of variance conducted on 5 German chamomile populations gathered from various regions of the country, the variation between populations was found to be 3%, while the variation within populations was 97%. Furthermore, the variation between environments was 0%. These results indicate that the diversity within the populations is significantly greater than the diversity between them. In 2013, Ahmadi reported results that were consistent with the latest findings. They discovered that intra-group diversity was higher (58.98%) than between-group diversity (41.02%) while studying the genetic diversity of German chamomile in 21 Iranian populations and two European populations. They used ISSR molecular markers and various morphological characteristics to conduct molecular variance analysis [10][16].

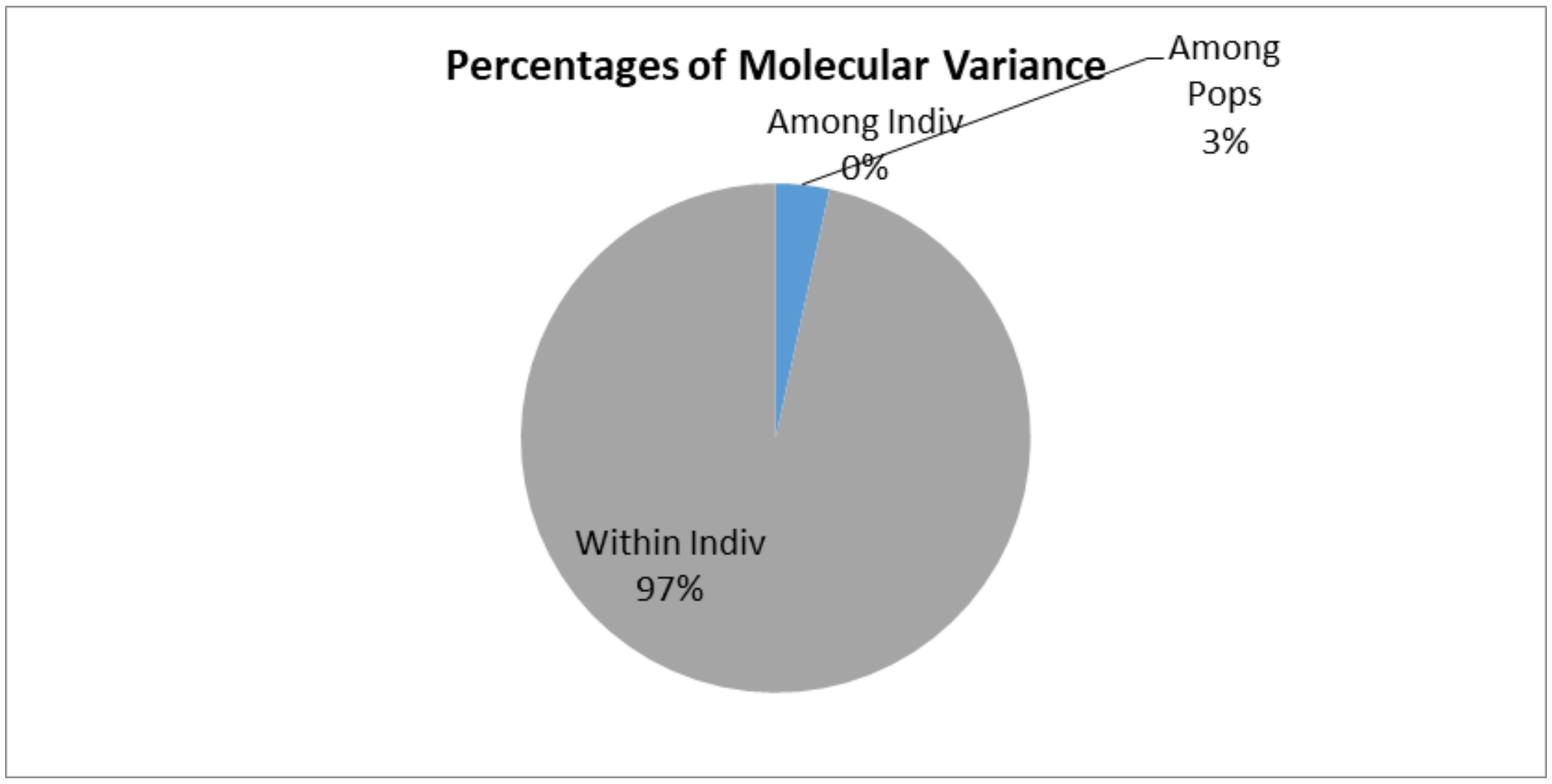

**Figure 2- Variation within and between chamomile populations based on SSR marker genotypic data in 5 German chamomile populations**

**Table 9- Analysis of molecular variance (AMOVA) of 5 German chamomile populations under study**

| Source of Variation | Degree of freedom | Sum of squares | Mean of squares | probability | Abundance percentage |
|---|---|---|---|---|---|
| Between environment | 4 | 5.300 | 1.325 | 0.052 | 3% |
| inter the population | 20 | 16.100 | 0.805 | 0.000 | 0% |
| inter the population | 25 | 37.000 | 1.480 | 1.480 | 97% |
| Total | 49 | 58.400 | - | 1.532 | 100% |

## Analysis of data obtained from SSR primers

By using the combination of 6 pairs of primers on the 5 studied populations, a total of 63 scorable bands were produced, of which 58 bands were polymorphic. The highest banding is related to population 2 (Saho Sarmak-Dashti) with 15 bands and the lowest banding is related to population 5 (Anarestan-Galobardakan) with 10 bands. The percentage of polymorphism was calculated between 66.67 and 100%. The highest percentage of polymorphism was related to Saho Sarmak-Dashti population. The anarestan-Galobaradakan population had the most monomorphic band and Saho Sarmak-Dashti population had the most polymorphic band. Dastmali et al. (2010) stated that a total of 1022 bands were observed from 13 pairs of selected primers, which showed 25.04% polymorphism in their studies on the evaluation of genetic diversity of 47 samples of Khatmi medicinal plants (Althaea & Alcea) using AFLP markers [6].

## The number of effective alleles, Shannon's index, Nei diversity coefficient, observed and expected heterozygosity

One of the important indicators for investigating genetic diversity among different cultivars and populations is the genetic diversity index of nei. (Nee, 1972). The estimation of nei index showed that the degree of genetic diversity of nei varied from 0.42 to 0.28. The KAM SSR-85 primer with a value of 0.42 had the highest and the KAM SSR-65 primer with a value of 0.28 had the lowest gene diversity among SSR markers. The average gene diversity for all primers used in the studied population was equal to Table 10-4. Using the Shannon coefficient, it is possible to show the amount of polymorphism among different genotypes. In this research, the Shannon coefficient in different primers varied between 0.69 and 0.41. KAM SSR-85 primer had the highest Shannon index with a value of 0.69 and KAM SSR-65 primer had the lowest Shannon index with a value of 0.41 among SSR markers. The number of effective alleles in the present study ranged from 1.520 to 1.838 with an average of 1.668. Anarestan-Galobardakan population had the lowest and Saho Sarmak-Dashti population had the highest number of effective alleles. The lowest and highest value of Shannon's index and nei diversity coefficient, as well as the number of effective alleles, were obtained in Anarestan-Galobardakan and Saho Sarmak-Dashti populations, respectively.
In 2011, Ramzanpour investigated the genetic diversity of 31 chamomile populations using AFLP markers and reported that the number of effective alleles ranged from 1.28 to 1.68 for E41-M60 and E35-M61 markers, respectively. The calculation of the genetic diversity and Shannon index also showed that the E41-M60 marker had the lowest diversity and Shannon index (0.2 and 0.34 respectively), while the highest amount of genetic diversity and Shannon index was obtained for the E35-M93 marker [7]. The observed heterozygosity in the present study was obtained in the range of 0.433 to 0.567. Also, the expected heterozygosity was calculated in the range of 0.287 to 0.423, which indicated the greatest genetic diversity in the Sahosarmak-Dashti population and the greatest diversity in the Anarestan-Galobardakan population.

## 2D plot of genotypes using principal coordinate decomposition

In this study, the populations of Tang Faryab and Anarestan Galobardakan are located next to each other in the main coordinate analysis grouping. The populations of Qalat Nilo and Malek Nowruz garden were also placed in separate groups in the graph obtained by analyzing the main coordinates. Also, the population of Saho Sarmak Dashti was placed in a separate grouping compared to other populations (Figure 7-4). The distribution of genotypes on the level of the diagram shows the appropriate diversity of the studied genotypes. Ramadanpour (2011) in a study using AFLP markers on 31 German chamomile populations and also Ahmadi in 2016 by evaluating the genetic diversity of 18 German chamomile populations, using AFLP markers, reported that the grouping resulting from cluster analysis and decomposition to original coordinates consistent with the original coordinates [7].

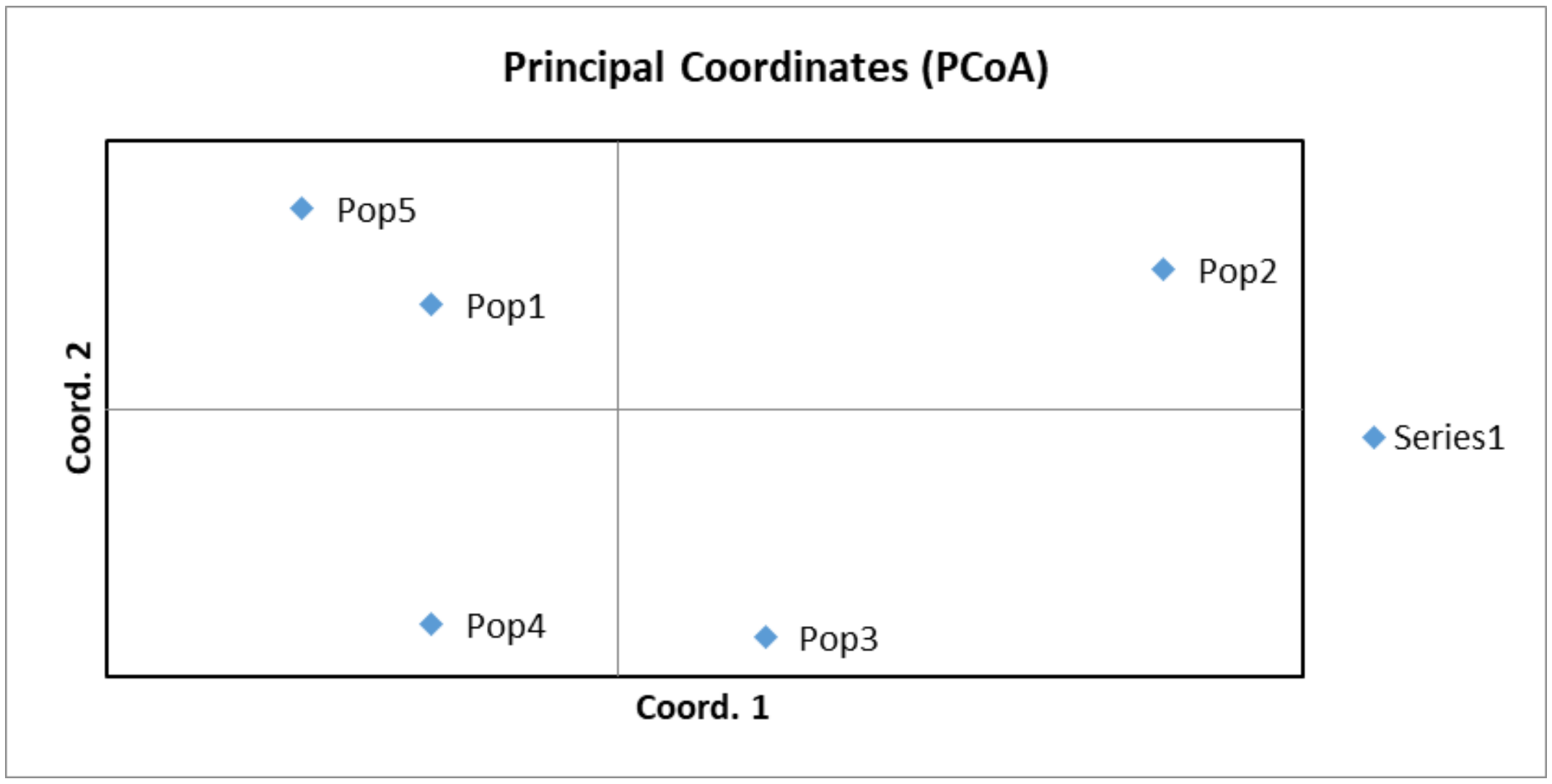

**Figure 3- The relationship between the examined genotypes using principal component analysis by GenALExe software**.

1 - Tang Faryab, 2- Saho Sarmak - Dashti, 3- Qalat Nilo, 4- Malek Nowruz Garden, 5- Anarestan - Galobardakan

Table 2- Simple variance analysis of the evaluated traits in German chamomile medicinal plant populations

(characteristics examined by MS mean square)

| Variation source | Degree of freedom | Day to greening | Day to flowering | Height | Flower diameter | Receptacle diameter | Fresh weight of flower | Dry weight of flower | Chlorophyll a | Chlorophyll b | Total Chlorophyll | Chamazulene % | Essential oil % |
|---|---|---|---|---|---|---|---|---|---|---|---|---|---|
| block | 2 | 1.28 ns | 1.05 ns | 21.36 ns | 0.623 ns | 0.01 ns | 11.168 | 0.506 ns | 0.0004 ns | 0.00011 ns | 0.0001 ns | 0.0012 ns | 0.33 ns |
| Treatment | 41 | 5.90 ns | 54.50 ** | 195.54 ** | 3.37 ** | 1.925 ** | 18.3531** | 1159.02 ** | 0.0072 ** | 0.0027 ** | 0.0027 ** | 0.023 ** | 17.65 ** |
| Experiment error | 82 | 1.20 | 14.77 | 42.78 | 1.23 | 0.377 | 85.39 | 18.84 | 0.0003 | 0.0004 | 0.0006 | 0.0007 | 0.31 |
| Coefficient of variation% | - | 5.65 | 5.47 | 11.61 | 5.69 | 9.06 | 7.9 | 9.67 | 11.72 | 14.62 | 17.50 | 13.33 | 9.59 |

Ns absence of significant difference, and ** significance at the statistical level of 1%.

Table 3- Average comparison results of the evaluated traits in different German chamomile populations

| Populations | Day to flowering | Height (cm) | Flower diameter (m4m) | receptacles diameter (mm) | fresh Flower weight per harvest (g). | Dry Flower weight per harvest (g). | Chlorophyll a (mg g-۱ FLW) | Chlorophyll b (mg g$^{-}$ $^{1}$FLW) | Total Chlorophyll | chamazulene (%) | Essential oil (%) |
|---|---|---|---|---|---|---|---|---|---|---|---|
| Tange faryab | 68.66 $^{b-e}$ | 53.66 $^{e-m}$ | 20.45 $^{a-d}$ | 6.28 $^{b-h}$ | 123.36$^{ghi}$ | 57.24 $^{ef}$ | 0.26$^{a}$ | 0.20 $^{a}$ | 0.20 $^{ab}$ | 0.35 $^{klm}$ | 0.13 $^{kmn}$ |
| Jam 2 | 73.33$^{bc}$ | 60.74$^{c-j}$ | 20.45 $^{a-d}$ | 6.26 $^{b-h}$ | 171.30 $^{ab}$ | 96.21$^{a}$ | 0.24 $^{abc}$ | 0.19 $^{ab}$ | 0.20 $^{abc}$ | 6.35 $^{ij}$ | 0.14 $^{k-n}$ |
| Izeh khozestan | 70.00 $^{b-e}$ | 55.54 $^{d-m}$ | 20.41 $^{a-d}$ | 6.38 $^{b-g}$ | 140.75 | 45.93 $^{hi}$ | 0.10 $^{no}$ | 0.12$^{l}$ | 0.11 $^{i-m}$ | 4.33 $^{no}$ | 0.13 $^{k}$-n |
| Dehrod sofla | 60.00 $^{f}$ | 73.83 $^{ab}$ | 17.55 $^{g}$ | 4.82 $^{ijk}$ | 168.86 $^{bc}$ | 65.79 $^{d}$ | 0.11 $^{mno}$ | 0.09 $^{l}$ | 0.11 $^{j-m}$ | 5.42 $^{klm}$ | 0.13 $^{lmn}$ |
| shahijan | 74.00 $^{bc}$ | 74.35$^{a}$ | 21.43 $^{b}$ | 6.90 $^{abc}$ | 180.20$^{ab}$ | 98.81 $^{a}$ | 0.22 $^{cd}$ | 0.14 $^{d-h}$ | 0.15 $^{d-i}$ | 10.37 $^{a}$ | 0.38 $^{ab}$ |
| lalab | 69.66 $^{b-e}$ | 46.66lmn | 19.54 $^{c-f}$ | 6.54 $^{b-g}$ | 82.12 $^{o-s}$ | 33.69 $^{k-n}$ | 0.14 $^{i-m}$ | 0.12 $^{g-k}$ | 0.13 $^{e-l}$ | 6.11 $^{kj}$ | 0.15 $^{i-m}$ |
| Haris | 70.66 $^{b-e}$ | 50.93 $^{j-n}$ | 19.45 $^{c-f}$ | 6.53 $^{b-g}$ | 122.98$^{hij}$ | 45.16 $^{hi}$ | 0.11 $^{mno}$ | 0.15 $^{d-h}$ | 0.14 $^{d-j}$ | 8.28 $^{efg}$ | 0.18 $^{h-k}$ |
| samosarmak | 69.00 $^{b-e}$ | 43.76 $^{n}$ | 18.55 $^{efg}$ | 6.33 $^{b-g}$ | 150.31 $^{de}$ | 64.94 $^{gh}$ | 0.17 $^{e-h}$ | 0.15 $^{d-g}$ | 0.17 $^{b-f}$ | 7.36 $^{h}$ | 0.13 $^{lmn}$ |
| urea | 66.66 $^{de}$ | 56.32 $^{c-l}$ | 15.52 $^{h}$ | 3.56 $^{l}$ | 74.57 $^{s}$ | 26.06 $^{p}$ | 0.09 $^{o}$ | 0.09 $^{l}$ | 0.09 $^{m}$ | 1.88 $^{s}$ | 0.11 $^{n}$ |
| Qalat nilo | ۷۳/۶۶ $^{bc}$ | 56.55 $^{c-l}$ | 21.56 $^{a}$ | 6.44 $^{b-g}$ | 182.17 $^{a}$ | 66.00 $^{d}$ | 0.18 $^{efg}$ | 0.15 $^{d-h}$ | 0.14 $^{e-l}$ | 9.91 $^{ab}$ | 0.14 $^{kn}$ |
| Malek Nowruz gardn | 74.00 $^{bc}$ | 55.90 $^{d-m}$ | 18.83 $^{c-g}$ | 6.23 $^{b-h}$ | 106.83 $^{lm}$ | 45.32 $^{hi}$ | 0.12$^{k-n}$ | 0.12 $^{h-k}$ | 0.13 $^{g-m}$ | 9.43 $^{bc}$ | 0.29 $^{de}$ |
| Baharestan jam | 70.33 $^{b-e}$ | 45.31$^{mn}$ | 20.09 $^{a-e}$ | 6.44 $^{b-g}$ | 135.29 $^{fgh}$ | 46.05 $^{ghi}$ | 0.14 $^{h-l}$ | 0.14 $^{d-i}$ | 0.14 $^{e-k}$ | 9.28 $^{bcd}$ | 0.38 $^{ab}$ |
| Asad abad behbahan | 68.66$^{b-e}$ | 52.80 $^{f-n}$ | 20.07$^{a-e}$ | 6.33 $^{b-g}$ | 113.09 $^{jkl}$ | 39.66$^{jik}$ | 0.14 $^{h-l}$ | 0.11 $^{jkl}$ | 0.13 $^{g-m}$ | 3.61 $^{op}$ | 0.15 $^{j-n}$ |
| Loran khozestan | 65.33$^{e}$ | 56.12 $^{d-l}$ | 18.69 $^{d-g}$ | 6.29 $^{b-h}$ | 89.59 $^{n-q}$ | 27.42 $^{op}$ | 0.20 $^{de}$ | 0.16 $^{cde}$ | 0.16 $^{b-g}$ | 7.25 $^{hi}$ | 0.34$^{bc}$ |
| Eslam abad jam | 71.00 $^{b-e}$ | 66.86 $^{abc}$ | 19.51$^{c-f}$ | 6.70 $^{a-d}$ | 155.81 $^{cd}$ | 53.04 $^{fg}$ | 0.15 $^{h-k}$ | 0.13$^{d-j}$ | 0.14 $^{e-l}$ | 8.43 $^{def}$ | 0.24$^{fg}$ |
| pashto | 70.00 $^{b-e}$ | 49.00 $^{k-n}$ | 19.40 $^{c-f}$ | 6.32 $^{b-g}$ | 138.93 $^{efg}$ | 40.76 $^{hij}$ | 0.12 $^{k-o}$ | 0.10 $^{kl}$ | 0.10 $^{lm}$ | 6.14 $^{jk}$ | 0.17 $^{h-l}$ |
| Samosarmak 2 | 69.66 $^{b-e}$ | 51.93 $^{h-n}$ | 19.12 $^{c-g}$ | 6.51 $^{b-g}$ | 82.47 $^{o-s}$ | 30.28 $^{l-p}$ | 0.16 $^{g-j}$ | 0.14 $^{d-h}$ | 0.14 $^{e-l}$ | 5.59 $^{jkl}$ | 0.33$^{cd}$ |
| galobardakan | 74.33 $^{b}$ | 55.44 $^{d-m}$ | 17.58 $^{g}$ | 4.22 $^{kl}$ | 87.70 $^{p-r}$ | 29.07 $^{m-p}$ | 0.24 $^{abc}$ | 0.20 $^{a}$ | 0.20 $^{ab}$ | 8.37 $^{ef}$ | 0.15 $^{j-n}$ |
| chamkalati | 65.00 $^{e}$ | 61.60 $^{c-i}$ | 19.93 $^{a-e}$ | 6.60 $^{b-f}$ | 80.40 $^{p-s}$ | 27.96 $^{op}$ | 0.19 $^{def}$ | 0.16 $^{c-f}$ | 0.16 $^{c-h}$ | 7.46 $^{gh}$ | 0.39 $^{a}$ |
| Kazeron | 67.00 $^{de}$ | 66.86 $^{abc}$ | 19.57 $^{c-f}$ | 6.72 $^{a-d}$ | 81.11 $^{nop}$ | 35.96 $^{j-m}$ | 0.11 $^{l-o}$ | 0.11$^{i-l}$ | 0.11 $^{j-m}$ | 8.74 $^{cde}$ | 0.36$^{abc}$ |
| Khoviz mountain | 71.66 $^{bcd}$ | 62.98 $^{c-f}$ | 20.12 $^{a-e}$ | 5.56 $^{ghi}$ | 173.87 $^{ab}$ | 81.25 $^{b}$ | 0.25 $^{ab}$ | 0.14 $^{d-i}$ | 0.14 $^{e-k}$ | 5.40 $^{klm}$ | 0.17 $^{h-l}$ |

Table 3- Average comparison results of the evaluated traits in different German chamomile populations

| population | Day to flowering | Height (cm) | Flower diameter (m4m) | Capitol diameter (mm) | Flower wet weight per harvesting (g) | Flower dry weight per harvesting (g) | Cholorophyll a (mg $g^{-1}$ FLW) | Cholorophyll b (mg $g^{-1}$FLW) | Total cholorophyll | chamazulene (%) | Essential oil (%) |
|---|---|---|---|---|---|---|---|---|---|---|---|
| Tange eram | 69.00 $^{b-e}$ | 61.52$^{c-j}$ | 19.24 $^{c-g}$ | 6.94 $^{ab}$ | 85.16 $^{o-s}$ | 28.25 $^{nop}$ | 0.16 $^{g-j}$ | 0.12 $^{h-l}$ | 0.12 $^{h-m}$ | 5.20 $^{lmn}$ | 0.24 $^{fg}$ |
| Token abdan | 70.33 $^{b-e}$ | 64.99 $^{a-d}$ | 19.40 $^{c-f}$ | 6.17$^{-h}$ | 121.52 $^{ijk}$ | 45.30 $^{hi}$ | 0.13 $^{j-n}$ | 0.11$^{i-l}$ | 0.11 $^{i-m}$ | 4.96 $^{lmn}$ | 0.12 $^{mn}$ |
| Jam 1 | 68.00 $^{cde}$ | 51.25$^{i-n}$ | 18.84$^{c-g}$ | 5.93 $^{c-h}$ | 178.71 $^{ab}$ | 70.10 $^{be}$ | 0.22 $^{cd}$ | 0.16 $^{cde}$ | 0.16 $^{c-h}$ | 9.9 $^{ab}$ | 0.40 $^{a}$ |
| Ghalea toil khozestan | 71.00 $^{b-e}$ | 52.03 $^{g-n}$ | 19.76 $^{a-f}$ | 6.71 $^{a-d}$ | 102.13 $^{lmn}$ | 27.01 $^{op}$ | 0.12 $^{k-o}$ | 0.11 $^{jkl}$ | 0.11 $^{j-m}$ | 2.97 $^{pqr}$ | 0.17 $^{h-m}$ |
| Tal bardi | 70.66 $^{b-e}$ | 63.69 $^{b-e}$ | 20.27 $^{a-e}$ | 6.35 $^{b-g}$ | 120.45 $^{ijk}$ | 33.64$^{k-o}$ | 0.10 $^{no}$ | 0.12 $^{h-l}$ | 0.15 $^{d-j}$ | 2.63 $^{qrs}$ | 0.15 $^{i-m}$ |
| East lavar | 70.66 $^{b-e}$ | 61.34 $^{c-j}$ | 19.26 $^{c-g}$ | 6.11 $^{b-h}$ | 123.27 $^{hij}$ | 35.12 $^{j-n}$ | 0.12 $^{k-n}$ | 0.11 $^{jkl}$ | 0.11 $^{i-m}$ | 5.43$^{klm}$ | 0.13 $^{lmn}$ |
| Degrode sofla 2 | 60.00 $^{f}$ | 52.66 $^{g-n}$ | 18.07$^{fg}$ | 4.40 $^{jkl}$ | 108.63 $^{kl}$ | 30.33 $^{l-p}$ | 0.13 $^{j-n}$ | 0.13 $^{e-j}$ | 0.13 $^{g-m}$ | 7.60 $^{fgh}$ | 0.16 $^{h-m}$ |
| unknown | 70.33 $^{b-e}$ | 58.66 $^{c-k}$ | 20.01 $^{a-e}$ | 6.68 $^{a-e}$ | 125.26 $^{hij}$ | 37.09 $^{jkl}$ | 0.18 $^{efg}$ | 0.19 $^{abc}$ | 0.18 $^{a-d}$ | 2.86 $^{pqr}$ | 0.13 $^{lmn}$ |
| Pol dokhtari | 71.00 $^{b-e}$ | 49.18 $^{k-n}$ | 19.78 $^{a-f}$ | 6.55 $^{b-g}$ | 109.25 $^{kl}$ | 47.58 $^{gh}$ | 0.11$^{l-o}$ | 0.15 $^{d-h}$ | 0.15 $^{d-j}$ | 2.75 $^{p-s}$ | 0.13 $^{lmn}$ |
| Jam and riz | 60.00 $^{f}$ | 62.56 $^{c-g}$ | 19.17 $^{c-g}$ | 6.70 $^{a-d}$ | 75.16 $^{rs}$ | 27.43 $^{op}$ | 0.17 $^{f-i}$ | 0.13 $^{f-k}$ | 0.12 $^{g-m}$ | 9.41 $^{bc}$ | 0.36$^{abc}$ |
| Zeraei Tehran | 73.33 $^{bc}$ | 56.49 $^{c-l}$ | 19.59 $^{c-f}$ | 5.90 $^{d-h}$ | 77.87 $^{qrs}$ | 28.21 $^{nop}$ | 0.11 $^{mno}$ | 0.11 $^{i-l}$ | 0.11$^{i-m}$ | 2.23 $^{rs}$ | 0.25 $^{ef}$ |
| Bona | 74.00 $^{bc}$ | 75.54 $^{a}$ | 20.24 $^{a-d}$ | 6.54 $^{b-g}$ | 148.18 $^{def}$ | 73.92 $^{c}$ | 0.17 $^{f-i}$ | 0.13 $^{f-k}$ | 0.12 $^{g-m}$ | 5.66 $^{jkl}$ | 0.14 $^{k-n}$ |
| Hamedan | 74.00 $^{bc}$ | 63.66 $^{b-e}$ | 20.58 $^{abc}$ | 7.67 $^{a}$ | 169.01 $^{b}$ | 74.52 $^{bc}$ | 0.15 $^{h-k}$ | 0.11 $^{jkl}$ | 0.10 $^{klm}$ | 3.51 $^{opq}$ | 0.12 $^{mn}$ |
| Harami anari | 73.66 $^{bc}$ | 74.71$^{a}$ | 19.78 $^{a-f}$ | 5.70 $^{e-i}$ | 86.08 $^{o-s}$ | 28.76 $^{nop}$ | 0.12 $^{k-o}$ | 0.13 $^{d-j}$ | 0.13 $^{e-l}$ | 5.41$^{klm}$ | 0.21 $^{fgh}$ |
| Kori manzar | 72.33 $^{bcd}$ | 54.17 $^{e-m}$ | 18.79 $^{c-g}$ | 6.09 $^{b-h}$ | 73.88 $^{s}$ | 26.97 $^{op}$ | 0.14 $^{i-m}$ | 0.11 $^{i-l}$ | 0.11 $^{j-m}$ | 3.49 $^{opq}$ | 0.16 $^{i-m}$ |
| Kalameh dashtestan | 72.66 $^{bc}$ | 61.84 $^{c-i}$ | 20.02 $^{a-e}$ | 6.56 $^{b-e}$ | 92.60 $^{nop}$ | 33.08 $^{k-p}$ | 0.16 $^{g-j}$ | 0.13 $^{d-j}$ | 0.15 $^{d-i}$ | 5.08 $^{lmn}$ | 0.15 $^{i-m}$ |
| Daneshkadeh Selected | 68.33 $^{b-e}$ | 58.15 $^{c-j}$ | 19.78 $^{a-f}$ | 5.65 $^{f-i}$ | 134.65 $^{gh}$ | 60.38 $^{de}$ | 0.17 $^{f-i}$ | 0.13 $^{d-j}$ | 0.13 $^{e-l}$ | 5.77 $^{jkl}$ | 0.20 $^{ghi}$ |
| Aale khorshid | 69.33 $^{b-e}$ | 62.15 $^{c-h}$ | 19.68 $^{b-f}$ | 5.31 $^{hij}$ | 85.34 $^{o-s}$ | 32.43 $^{l-p}$ | 0.12 $^{k-n}$ | 0.13 $^{d-j}$ | 0.12 $^{h-m}$ | 4.62 $^{mn}$ | 0.19 $^{hij}$ |
| Galobardakan (pankareh) | 73.33 $^{bc}$ | 63.63 $^{b-e}$ | 18.59 $^{efg}$ | 4.46 $^{jkl}$ | 80.44 $^{p-s}$ | 32.48 $^{l-p}$ | 0.24 $^{abc}$ | 0.16 $^{de}$ | 0.15 $^{d-j}$ | 2.81 $^{prs}$ | 0.25 $^{ef}$ |
| Tekyeh gard (khozestan) | 71.66 $^{bcd}$ | 49.11 $^{k-n}$ | 20.21 $^{a-e}$ | 6.25 $^{b-h}$ | 129.18 $^{ghi}$ | 35.87 $^{j-m}$ | 0.22 $^{bcd}$ | 0.16$^{bdc}$ | 0.17 $^{a-e}$ | 3.23 $^{pq}$ | 0.19 $^{hij}$ |
| Abnormal receptacles | 84.00 $^{a}$ | 46.64 $^{lmn}$ | 20.50$^{abc}$ | 5.64 $^{f-i}$ | 93.94 $^{mno}$ | 28.03 $^{op}$ | 0.26 $^{a}$ | 0.20$^{a}$ | 0.21 $^{a}$ | 3.10 $^{pqr}$ | 0.17 $^{h-l}$ |

Table 5- Simple phenotypic correlation coefficients between the evaluated traits in different German chamomile populations

| Characteristics | Essential oil % | Day to greening | Day to flowering | Height | Flower diameter | Receptacles diameter | Fresh weight of flower | Dry weight of flower | Cholorophyll a | Cholorophyll b | Total cholorophyll | Chamazulene % |
|---|---|---|---|---|---|---|---|---|---|---|---|---|
| Essential oil % | 1 | | | | | | | | | | | |
| Day to greening | 0.007 | 1 | | | | | | | | | | |
| Day to flowering | z | 0.474** | 1 | | | | | | | | | |
| Height | 0.070 | 0.016 | 0.064 | 1 | | | | | | | | |
| Flower diameter | 0.217 | 0.130 | 0.150 | 0.009 | 1 | | | | | | | |
| Receptacles diameter | -0.017 | 0.113 | 0.440** | 0.057 | 0.672** | 1 | | | | | | |
| Fresh weight of flower | -0.143 | 0.031 | 0.082 | 0.154 | 0.239 | 0.430** | 1 | | | | | |
| Dry weight of flower | -0.016 | 0.091 | 0.137 | 0.310* | 0.212 | 0.412** | 0.882** | 1 | | | | |
| Cholorophyll a | 0.158 | -0.025 | 0.301* | -0.073 | -0.050 | 0.265 | 0.222 | 0.332* | 1 | | | |
| Cholorophyll b | 0.074 | -0.051 | 0.347* | -0.222 | -0.012 | 0.253 | 0.089 | 0.140 | 0.801** | 1 | | |
| Total cholorophyll | 0.042 | -0.034 | 0.309* | 0.243 | 0.019 | 0.284* | 0.160 | 0.190 | 0.798** | 0.952** | 1 | |
| Chamazulene % | 0.565** | 0.052 | -0.220 | 0.061 | 0.168 | 0.011 | 0.317* | 0.336* | 0.163 | 0.163 | 0.157 | 1 |

* and ** significance at the statistical level of 5% and 1% respectively.

Table 6: Step-by-step regression analysis of the essential oil of different chamomile populations as a dependent variable and other traits as an independent variable

| Characteristic in the order of entry in the model | Code of independent Characteristics entered into the model | Regression coefficient of traits | Coefficient of explanation | Cumulative coefficient |
|---|---|---|---|---|
| Chamazulene % | 11X | 0.024 | 0.319 | 0.319 |
| Fresh weight of flower | 6X | -0.000093 | 0.116 | 0.436 |
| Intercept | 0.172 | | | |
| | | | | $Y = -0.000093x_6 + 0.024x_{11} + 0.172$ |

Table 7- Causality analysis results for essential oil percentage in 42 German chamomile populations

| Characteristic | Direct effects | Indirect effects through | | Correlation with essential oil percentage |
|---|---|---|---|---|
| | | chamazolene | Fresh weight | |
| Chamazolene | 0.608 | - | -0.120 | $0.56^{**}$ |
| Fresh weight | -0.360 | 0.343 | - | -0.017 |

* and ** are significant at the five and one percent probability level, respectively

Table 8-: principle components analysis for all evaluated traits in the 42 investigated German chamomile populations

| characteristics | Component 1 | Component 2 | Component 3 | Component 4 | Component 5 |
|---|---|---|---|---|---|
| Day to greening | 0.075 | 0.112 | -0.307 | 0.351 | 0.659 |
| Day to flowering | 0.259 | -0.077 | -0.427 | 0.171 | 0.275 |
| Height | -0.023 | 0.338 | 0.065 | -0.202 | 0.304 |
| Flower diameter | 0.180 | 0.333 | -0.101 | 0.478 | -0.432 |
| Diameter of Flower Receptacles | 0.334 | 0.244 | -0.264 | 0.200 | -0.358 |
| Fresh weight of flower | 0.303 | 0.403 | 0.016 | -0.362 | -0.033 |
| Dry weight of flower | 0.332 | 0.400 | 0.055 | -0.335 | 0.118 |
| Chlorophyll a | 0.425 | -0.259 | 0.125 | -0.107 | 0.095 |
| Chlorophyll b | 0.417 | -0.348 | 0.073 | -0.042 | -0.043 |
| Total chlorophyll | 0.429 | -0.348 | -0.073 | -0.042 | -0.043 |
| Chamazulene % | 0.057 | 0.058 | 0.508 | 0.493 | 0.119 |
| Essential oil % | 0.171 | 0.195 | 0.531 | 0.188 | 0.202 |

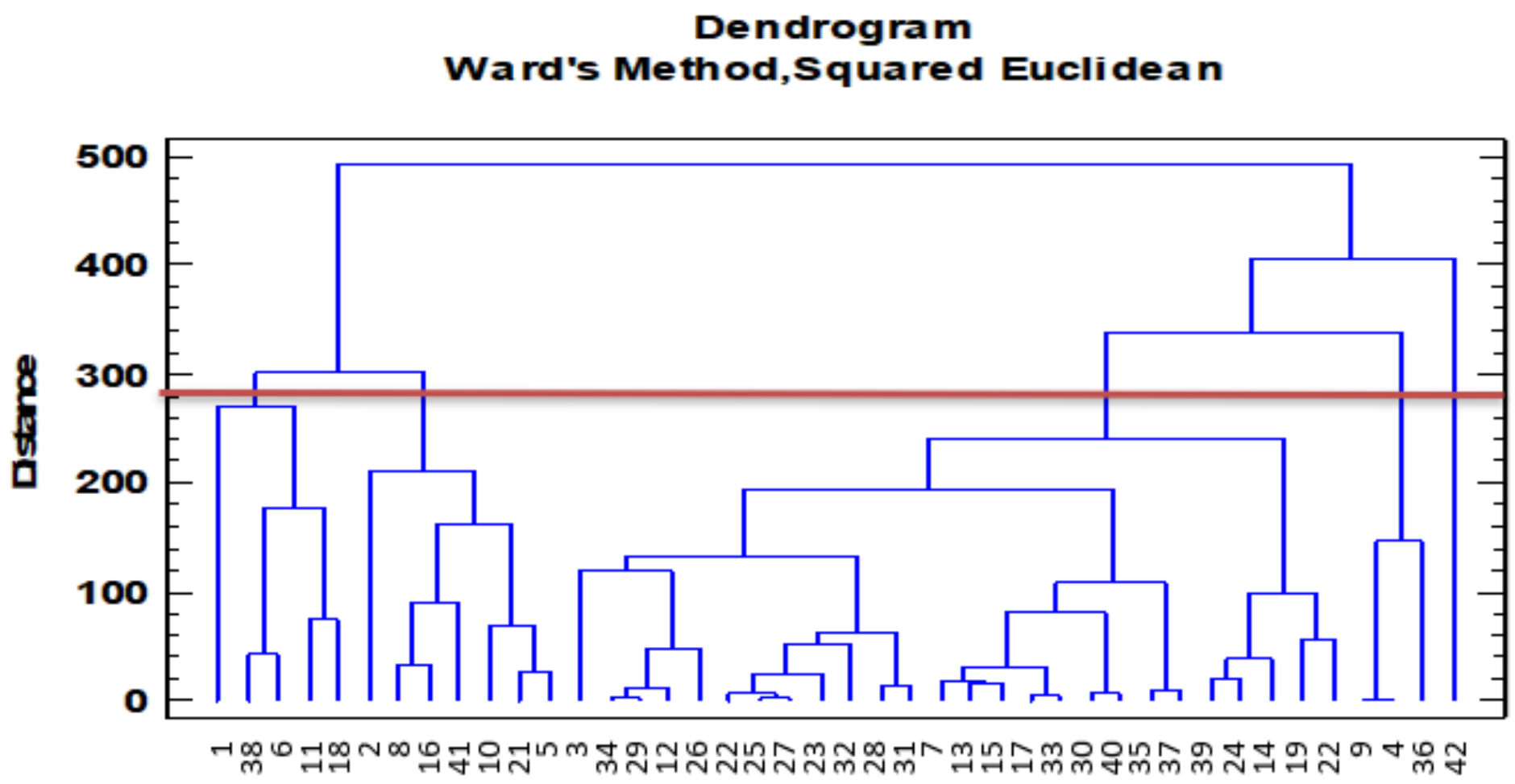

**Figure 2-4: Dendrogram resulting from the cluster analysis of 42 German chamomile populations based on the studied morphological and phytochemical .traits**

1- Tang Faryab. 2- Jam2. 3- Izeh Khuzestan. 4- Dehrud Sofla. 5- Shahijan, Fars province. 6- Lalab Khuzestan. 7- Harris village. 8- Sahosarmak dashti. 9- Urea chamomile. 10- Qalat Nilo. 11- Malek Nowruz garden. 12- Baharestan Jam. 13- Asadabad Behbahan. 14- Loran, Khuzestan. 15- Islamabad Jam. 16- Pashto 17- Samosarmak 2. 18- Anarestan. 19- Chamkalati Chamomile. 20- Kazeron. 21- Khoviz mountains. 22- Tang Eram 23- Noken Abdan. 24- Gam 1. 25- Ghalea Toli. 26- Tal Baradi. 27- Laversharghi 28- Dehrod sofla 2. 29- Unknown. 30- Pol Dokhtari. 31- Jam and Riz. 32- Zaraei Tehran. 33- Buna 34- Hamedan 1. 35- Harami anari Chamomile. 36- Kori manzar. 37- Kalameh Dashestan. 38- Elective of the Faculty of Agriculture in Borazjan. 39- Al Khursheed. 40- Galobardakan. 41- Tekyeh gard. 42- German Chamomile with abnormal receptacles.

Table 10- The results of applying 6 SSR primer combinations to 5 German chamomile populations

| Row | Population 1 | Population 2 | Population 3 | Population 4 | Population 5 | Summation | Average |
|---|---|---|---|---|---|---|---|
| Total bands | 13 | 15 | 13 | 12 | 10 | 63 | 12.6 |
| Number of monomorphic bands | 1 | 0 | 1 | 1 | 2 | 5 | 1 |
| Number of polymorphic bands | 12 | 15 | 12 | 11 | 8 | 58 | 11.6 |
| Polymorphic percentage | 83.33 | 100 | 83.33 | 83.33 | 66.67 | - | 83.33 |

**Table 11- Investigating the number of effective alleles, Shannon's index and reed diversity coefficient in 5 populations of German chamomile using SSR markers**

| Population | | Shannon Index | | | |
|---|---|---|---|---|---|
| Tange faryab | 1.770 | 0.585 | 0.377 | 0.533 | **0.377** |
| Saho sarmak, Dashti | 1.838 | 0.692 | 0.423 | 0.567 | **0.423** |
| Qalat Nilo | 1.637 | 0.565 | 0.357 | 0.433 | **0.357** |
| Malek Nouroz Garden | 1.574 | 0.504 | 0.327 | 0.467 | **0.327** |
| Anarestan, Galobardakan | 1.520 | 0.413 | 0.287 | 0.467 | **0.287** |

## Conclusion

According to the obtained results, Jam 1 and Shahijan populations had the most effective substances (essential oil and chamazulene). The correlation analysis revealed a positive and significant correlation of 59% between the percentage of chamazulene and the percentage of essential oil. Chamazulene percentage and fresh weight as the most important traits were entered into the regression model step by step. These traits were found to explain 0.43% of the changes in the data. These findings have significant implications for future research aimed at identifying the most effective populations for essential oil and chamazulene production. The cluster analysis divided the genotypes into five groups. The second group was the most important, and Jam1 and Shahijan genotypes were in this group. The results of the molecular analysis showed that the Seho Sermak-Dashti population had the most effective allele, and this population was superior to other populations in terms of Shannon's index and nei diversity coefficient. According to the diagram, analysis into main coordinates showed that the genotypes are scattered on the surface of the diagram and this indicates the appropriate diversity of the studied genotypes. As a result, it can be stated that the grouping of phenotypic and molecular data was very consistent with each other. Furthermore, it seems that more investigations are required in order to quantitative genetics mechanisms of German chamomile genotypes.

**Acknowledgments:** We gratefully acknowledge the faculties, staffs and technicians of the Faculty of Agriculture, Persian Gulf University, Bushehr, Iran for the providing research field in the present investigation. Further, we declare our special thanks to the Presov University, Presov, Slovakia for their valuable support and cooperation.

**Author Contributions:** M.M. and M.H.P. conceived and designed the experiments; M.H.P. performed the experiments; M.M., I.S., and M.G. and M.H.P. analyzed the data; M.G. and I.S. contributed reagents/materials/analysis tools; M.G. and M.M. and I.S wrote the paper.

**Conflicts of Interest:** The authors declare no conflict of interest.